\documentclass[11pt,a4paper]{article}
\pdfoutput=1
\usepackage{jcappub}
\usepackage{amsfonts}
\usepackage{amssymb}
\usepackage{epsfig}
\usepackage{ifpdf}
\usepackage{graphicx}
\usepackage{amsmath}

 \newlength{\wth}
\newcommand{\be}{\begin{equation}}
\newcommand{\ee}{\end{equation}}
\newcommand{\bea}{\begin{eqnarray}}
\newcommand{\eea}{\end{eqnarray}}

\newcommand{\ti}{\times}

\newcommand{\beqa}{\begin{eqnarray}}
\newcommand{\eeqa}{\end{eqnarray}}

\title{Constraints on Axion-Like Particles from Non-Observation of Spectral Modulations for X-ray Point Sources}
\author[1]{Joseph P. Conlon,}
\author[1]{Francesca Day,}
\author[1]{Nicholas Jennings,}
\author[1]{Sven Krippendorf,}
\author[2]{Markus Rummel}
\affiliation[1]{Rudolf Peierls Centre for Theoretical Physics, 1 Keble Road, Oxford, OX1 3NP, UK}
\affiliation[2]{Physics and Astronomy, McMaster University, Hamilton, ON, Canada, L8S 4M1}
\emailAdd{joseph.conlon@physics.ox.ac.uk}
\emailAdd{francesca.day@physics.ox.ac.uk}
\emailAdd{nicholas.jennings@physics.ox.ac.uk}
\emailAdd{sven.krippendorf@physics.ox.ac.uk}
\emailAdd{rummelm@mcmaster.ca}

\abstract{We extend previous searches for X-ray spectral modulations induced by ALP-photon conversion
to a variety of new sources, all consisting of quasars or AGNs located in or behind galaxy clusters.
We consider a total of seven new sources, with data drawn from the \emph{Chandra} archive. In all cases the spectrum is well fit by an absorbed
power-law with no evidence for spectral modulations, allowing constraints to be placed on the ALP-photon coupling parameter $g_{a\gamma\gamma}$.
Two sources are particularly good: the Seyfert galaxy 2E3140 in A1795 and the AGN NGC3862 within the cluster A1367, leading to 95\% bounds for light ALPs ($m_a \lesssim 10^{-12} {\rm eV}$) of
$g_{a\gamma\gamma} \lesssim 1.5 \ti 10^{-12} {\rm GeV}^{-1}$ and $g_{a\gamma\gamma} \lesssim 2.4 \ti 10^{-12} {\rm GeV}^{-1}$ respectively. }

\begin{document}
\maketitle

\section{Introduction}

Axion-like particles (ALPs) are one of the most promising extensions of the Standard Model. They
arise generically in string compactifications (e.g. \cite{hepth0602233, hepth0605206, 1206.0819}).
As their Lagrangian is protected by a shift symmetry,
ALPs can easily have masses that are extremely small or vanishing.
The canonical coupling of an ALP $a$ to electromagnetism is through the interaction
\be
\label{ALPphoton}
g_{a \gamma \gamma}~a~{\bf E} \cdot {\bf B}~,
\ee
where $g_{a\gamma\gamma}$ is the ALP-photon coupling
and ${\bf E}$ and ${\bf B}$ are the electric and magnetic fields. A general review of ALPs and their physics can be found in
\cite{RingwaldReview}.

If ALPs exist, then the interaction of equation (\ref{ALPphoton}) causes ALPs and photons to interconvert in the presence of a background magnetic field $\langle B \rangle$~\cite{Sikivie:1983ip, Raffelt}. This phenomenon is at the heart of most ALP searches, including the planned experiment IAXO \cite{IAXO}.
A variety of search strategies can be used depending on the ALP mass. In this paper
we concern ourselves with the case of light ALPs, $m_a \lesssim 10^{-12} {\rm eV}$.\footnote{$10^{-12} {\rm eV}$ is the plasma frequency of a galaxy cluster,
and conversion is suppressed for ALP masses above this value.}
In this case
non-observation of gamma ray photons coincident with the
SN1987A neutrino burst constrains $g_{a \gamma \gamma} \lesssim 5 \cdot 10^{-12} {\rm GeV}^{-1}$~\cite{Brockway:1996yr,Grifols:1996id,1410.3747}.

\begin{figure}
\centering
\includegraphics[width=0.49\textwidth]{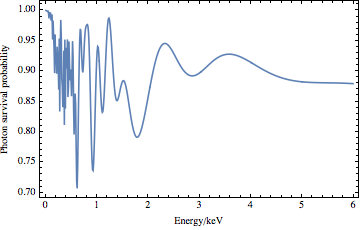} \includegraphics[width=0.49\textwidth]{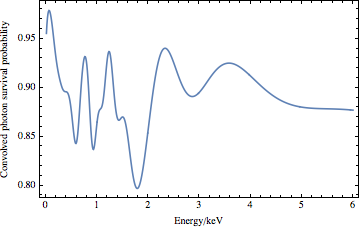}
\caption{Left---The photon survival probability for a photon propagating from NGC3862 in A1367 (see below) to us. The ALP-photon couping is $g_{a\gamma\gamma} = 2 \ti 10^{-12} {\rm GeV}^{-1}$. Note that the precise form of the survival probability depends on the unknown configuration of the magnetic field along the line of sight. However, the oscillatory structure is generic.
Right---The photon survival probability for the same magnetic field convolved with a Gaussian with FWHM of 150 eV, the energy resolution of {\it Chandra}.
 \label{FullALPPhotonConversion}}
\end{figure}

Galaxy clusters are particularly efficient photon-ALP converters~\cite{09014085, 0902.2320, WoutersBrun, 1305.3603, 1312.3947, 150101642, 150702855, 1509.06748}.
For the magnetic fields and electron densities present within galaxy clusters,
at X-ray energies
the $\gamma \leftrightarrow a$ conversion probability is both energy-dependent and
highly oscillatory (and even more so when polarisation is considered \cite{12046187, 150906748,  161205697}).
For a bright photon source located behind or in a cluster magnetic field,
the $\gamma \leftrightarrow a$ interconversion leads to modulations in the spectrum of
arriving photons. An example of the photon survival probability that would be imprinted on the spectrum is shown in Figure~\ref{FullALPPhotonConversion}. Although the survival probability as a function of photon energy depends on the precise configuration of the magnetic field along the line of sight, the oscillatory structure is generic. We also show the survival probability convolved with a Guassian with full width at half maximum 150 eV, representing the energy resolution of {\it Chandra}. This significantly smears out the ALP induced oscillations. Future missions, such as ATHENA, with substantially improved energy resolution will be sensitive to finer oscillations.
Compared to the source spectrum, this imprints oscillatory features on the
spectrum of arriving photons. A search
for such modulations can then lead to constraints on the coupling parameter $g_{a\gamma\gamma}$.

For this purpose, quasars or AGNs that are either behind or embedded in galaxy clusters provide attractive sources.
The original photon spectrum is reasonably well described by an absorbed power-law, and all the photons experience
a single sightline through the cluster (the kiloparsec scale on which cluster magnetic fields vary is far greater
than the size of the source). AGNs are also bright sources, enabling them to provide the large number of counts necessary for
statistical sensitivity to modulations of the photon spectrum.

Bright X-ray point sources also offer another potential physics application. If X-rays have a significant interaction rate with dark matter (for example,
an absorption resonance), then these interactions can also lead to spectral features~\cite{Profumo:2006im,160801684}.
In AGN and quasar spectra, the X-rays originate from within
a few Schwarschild radii of a supermassive central black hole. They therefore experience a dark matter column density
that is sensitive not just to the overall dark matter distribution of the host galaxy or galaxy cluster, but also to any dark matter spikes present on small scales all the way
down to milli-parsec distances from the central black hole.

For X-ray point sources, this method for constraining ALPs was first described and used in~\cite{WoutersBrun} (see also~\cite{1205.6428, BraxWoutersBrun} and \cite{1603.06978} for
a recent analysis of NGC1275 in gamma rays using this approach).
However,~\cite{WoutersBrun} only applied these ideas to the study of the AGN at the centre of the Hydra A galaxy cluster (redshift $z=0.052$), for which 240ks of
\emph{Chandra} observation time exist. The combination of the intrinsic AGN brightness and the
redshift of $z=0.052$ results in only
a few thousand counts in total, with a limited ability to discriminate against the cluster background.

A much better source than Hydra A is the central AGN of the Perseus cluster, at the heart of the galaxy NGC1275. This very bright source shines through the
Perseus cluster and is the subject of 1.5 Ms of \emph{Chandra} observation time (although all but 200ks  is highly contaminated by pileup).
The resulting spectrum has been studied in \cite{160501043, 160801684}. Based on a central magnetic field of $B_0 = 25 \mu {\rm G}$, the absence
of any spectral modulations beyond the 10\% level constrains $g_{a\gamma\gamma} \lesssim 1.5 \ti 10^{-12} {\rm GeV}^{-1}$.
There is also potential structure away from the best-fit power-law. In particular, there is a dip at
a best-fit energy of $(3.54 \pm 0.02) {\rm keV}$ -- the same energy as the diffuse excess of \cite{Bulbul, Boyarsky} -- and the consequences of such a dip
for the \emph{Hitomi} data and the 3.5 keV line have been explored in \cite{160801684}.

Recently, a similar analysis was carried out for the central AGN of the Virgo cluster at the heart of the galaxy M87 \cite{170307354}.
Although the AGN is not as bright as for NGC1275, the deep $ > 300{\rm ks}$ observation has a large number of counts.
Both M87 and NGC1275 are giant elliptical galaxies at the heart of large cool-core clusters, and comparable assumptions on the central
magnetic field (this time using $B_0 = 31 \mu {\rm G}$ for M87) led to a similar exclusion limit $g_{a\gamma\gamma} \lesssim 1.5 \ti 10^{-12} {\rm GeV}^{-1}$.

In this paper we extend these studies by  considering further X-ray point sources suitable for constraining ALP-photon mixing.
While not as individually bright as NGC1275 or M87, they all contain more counts than for
the Hydra A AGN. A further advantage of considering multiple sources is that the precise magnetic field along the
line of sight to any single source is always unknown. The use of multiple sources mitigates
the risk that a single source has a line of sight with a magnetic field configuration that is particularly unfavourable for ALP-photon conversion.

The paper is structured as follows. Section~\ref{sec:sources} describes the sources, data and methodology used to set bounds on the ALP-photon coupling parameter.
Section~\ref{sec:indsources} describes the results and bounds for each source, while Section~\ref{sec:conclusions} places these methods in the context of the future sensitivity of X-ray astronomy
to searches for ALPs.

\section{Sources and Methodology}
\label{sec:sources}

This paper is based on exploiting the substantial existing X-ray archives to search for sources that can lead to constraints on ALPs.
We have focused on the \emph{Chandra} archive for this purpose.
It is essential to the physics here that the sources shine through galaxy clusters.
This implies the presence of a contiguous background that has to be distinguished from the source.
\emph{Chandra} has by far the best angular resolution of X-ray telescopes, and
as for all except the very brightest sources good angular resolution is essential, singling out \emph{Chandra}
as the preferred instrument.

The sources used in this paper are found by examination of the \emph{Chandra} archive, using a combination
of manual inspection of images and SIMBAD \cite{SIMBAD} to locate bright point sources shining through galaxy clusters.
We have focused on nearby clusters, as they have a larger footprint on the sky. This implies that they
 are more likely to have a distant source behind them, and for the cases of sources embedded in the cluster also allows a greater discrimination of the point source
 from the contiguous diffuse emission.

As we rely on archival data,
the sources used here are not necessarily intrinsically optimal -- for any source we rely on the existence of sufficient observational time
to produce a usable number of photons. As one example, we use in this paper a $z=1.3$ quasar that shines through the cluster A1795. This
is usable solely because there is a total of 985ks of \emph{Chandra} time including it in the field of view. A much
shorter exposure would have contained insufficient photons to be useful.

\subsection{Enumeration of Sources Used}

We list here the seven point sources included within this paper. We exclude previously studied sources (Hydra A, NGC1275 and M87).
We assume a cosmology with $H_0 = 73 {\rm km s}^{-1}$, and
source redshifts have been identified using either
SIMBAD \cite{SIMBAD} or NED\footnote{http://ned.ipac.caltech.edu/}.

\begin{itemize}

\item The quasar B1256+281 at redshift $z=0.38$ at (RA,DEC) = (12:59:17,+27:53:46) shining through
the Coma cluster (z=0.023). This is offset by $521^{''}$ from the cluster centre, at a projected distance of 232kpc, and
there is 493ks of \emph{Chandra}
time containing it.

\item The quasar SDSS J130001.48+275120.6 at redshift $z=0.975$ at (RA,DEC) = (13:00:01, +27:51:20) shining through the
Coma cluster (z=0.023), offset by $484^{''}$ from the cluster centre at a projected distance of 215kpc, and
with 574ks of \emph{Chandra} time on it.

\item The bright AGN, NGC3862, within the cluster A1367 (z=0.0216) and at an offset of $443^{''}$ from the cluster centre (a projected distance of 186kpc),
at (RA,DEC) = (11:45:05, +19:36:23), and the subject of 75ks of
public \emph{Chandra} time.


\item The central AGN IC4374 of the cluster A3581 (z=0.023), at (RA, DEC)=(14:07:29, -27:01:04), with 85ks of \emph{Chandra} time.

\item The bright Sy1 galaxy 2E3140 at z=0.05893 within A1795 located at (RA,DEC) = (13:48:35, +26:31:09) and offset from the
center of the cluster by $417^{''}$ (a projected distance of 456kpc). In total, this is the subject of 660ks of \emph{Chandra} time.

\item The quasar CXOU J134905.8+263752 at $z=1.30$ behind A1795, at (RA, DEC) = (13:49:06, 26:37:48) and offset from the cluster center
by $177^{''}$ at a projected distance of 194kpc, and the subject of 985ks of \emph{Chandra} time.

\item The central AGN UGC9799 of the cluster A2052 (redshift z=0.0348) at (RA, DEC) = (15:16:45, +07:01:18), with 654ks of \emph{Chandra} time.


\end{itemize}

Data has been reduced and processed using CIAO 4.8.1. Different observations of the same source have all been stacked,
and backgrounds subtracted from a contiguous region.
 Although AGNs are variable, this variability does not affect the physics here. Any ALP-induced spectral deviation depends only on the line of sight,
and so is unaffected by variable strength of a source. The line of sight for each source is, for all practical purposes, identical.
This is because the cluster magnetic fields that source ALP-photon conversion are ordered on kiloparsec scales, while the emission in
AGNs arises on a scale comparable to the Schwarzschild radius of the black hole.

Pile-up is not a contaminant here, as these sources are intrinsically much less bright than either NGC1275 \cite{160501043} or
M87 \cite{170307354}. Furthermore, in many cases they have also been observed off-axis, as secondary background point sources within primary
observations of the host cluster, spreading out their images and further reducing pileup.

\subsection{Magnetic Field Models}

ALP-photon conversion requires a magnetic field. Direct measurements of magnetic fields of galaxy clusters are obtained via Faraday rotation and
require the presence of polarised radio sources either within or behind the cluster, as for the measurements of the
Coma magnetic field in \cite{Bonafede} or A2199 in \cite{12014119}. This is not possible
for all clusters, as the necessary radio sources may not exist. The recent paper on the A194 magnetic field \cite{A194} summarises extant measurements of cluster magnetic fields
(see its Table 8).

This paper involves both sources that are at the centre of a cluster, and also sources that are significantly offset from the center.
ALP-photon constraints depend on the magnetic field along the line of sight from the source. For sources whose projected position is at a significant
offset from the centre of the cluster, the field along the line of sight depends on \emph{both} the overall central magnetic field of the cluster, denoted $B_0$,
\emph{and} the rate at which the field decreases away from the centre.

This is parametrised by assuming the cluster magnetic field to be radially symmetric,
$$
B(r) \sim B_0 \left( \frac{n_e(r)}{n_0} \right)^{\eta}.
$$
$\eta$ is expected to lie somewhere between 0.5 and 1. The former represents
an equipartition between magnetic field energy and themal energy ($\langle B^2 \rangle \sim n_e k_B T$), while a value $\eta = 2/3$ corresponds to a magnetic field that is frozen into
the gas.
A value of $\eta = 1$ has been
found for the cool-core cluster Hydra A \cite{KucharEnsslin} (with a best-fit central magnetic field
$B_0 = 36\ \mu {\rm G}$). We use an intermediary value of $\eta = 0.7$.
A $\beta$-model for the electron density takes the form
$$
n_e(r) = n_0 \left( 1 + \frac{r^2}{r_c^2} \right)^{-3 \beta/2},
$$
where $r_c$ is the core radius. Although not perfect, the $\beta$ model captures the gross behaviour of the electron density in a cluster.
The parameters we use for each cluster are shown in Table~\ref{SourceMagneticFields}.
\begin{table}\begin{center}
\begin{tabular}{| c | c | c | c | c | c | c |}
\hline
Source & Cluster & $n_{e,0}$ & $r_c$ & $\beta$ & $B_0$ & $L_{total}$ \\
& & $(10^{-3} {\rm cm}^{-3})$ & (kpc) & & $\mu G$ &  (Mpc) \\
\hline
B1256+281 & Coma & 3.44 & 291 & 0.75 & 4.7  & 2 \\
\hline
SDSS J130001.47+275120.6 & Coma & 3.44 & 291 & 0.75 & 4.7  & 2 \\
\hline
NGC3862 & A1367 & 1.15 & 308 & 0.52 & 3.25 & 1 \\
\hline
IC4374 & A3581 & 20 & 75 & 0.6  & 1.5  & 1\\
\hline
2E3140 & A1795 & 50 & 146 & 0.631  & 20  & 1\\
\hline
CXOUJ134905.8+263752 & A1795 & 50 & 146 & 0.631  & 20  & 2 \\
\hline
UGC9799 & A2052 & 35 & 32 & 0.42 & 11 & 1 \\
\hline
\end{tabular}\end{center}
\caption{Parameters for the electron density and magnetic field models used for each of the clusters.
All sources use $\eta = 0.7$, $L_{min} = 1 {\rm kpc}$ and $L_{max} = 17 {\rm kpc}$. For Coma the parameters are taken from
\cite{Bonafede}. For A1367 the $\beta$-model parameters come from \cite{Ensslin97}
and the magnetic field from the article by M.~Henriksen in \cite{Henriksen}.
For A1795 the central magnetic field is taken from \cite{GeOwen1993}, the $\beta$-model parameters from \cite{Ettori2000} and the central
electron density from \cite{ChandranDennis2005}. For A2052 the parameters are taken from \cite{Blanton2003} and \cite{Machado2014} (correcting an error in the
conversion of the core radius from arcseconds to kiloparsecs). For the poor cluster A3581 the central electron density is taken from \cite{Johnstone98, Johnstone2005}. We could not
find beta model parameters in the literature and have used illustrative values of $r_c = 75{\rm kpc}$ and $\beta=0.6$. For the central magnetic field we
used the value for the poor cluster A194 \cite{A194} of $B_0 = 1.5 \mu{\rm G}$.}
\label{SourceMagneticFields}
\end{table}

The typical coherence lengths of the magnetic field are also relevant for ALP-photon conversion.
The magnetic field is expected to be multi-scale, with a power spectrum
that is distributed on a range of scales $P(L) \sim L^{-n}$ between some minimal and maximal length
$\Lambda_{min}$ and $\Lambda_{max}$. For the Coma magnetic field model of \cite{Bonafede}, minimal and maximal lengths were $\Lambda_{min} = 2 {\rm kpc}$
and $\Lambda_{max} = 34 {\rm kpc}$. In absence of other information, for other clusters
we have based the coherence lengths used on those in Coma.

We model the magnetic field as a series of 1-dimensional domains.
The lengths of the domains are drawn from a power-law distribution, restricted to a minimal and maximal length $L_{min}$
and $L_{max}$.\footnote{For the Coma field, we use $L_{min} = 0.5 \Lambda_{min}$ and $L_{max} = 0.5 \Lambda_{max}$. This is because
$\Lambda_{min}$ and $\Lambda_{max}$, as used in \cite{Bonafede}, are full 2$\pi$ periods, over which the magnetic field reverses orientation.}
The number of domains is chosen to match the total extent of the path through the cluster. For sources that are behind a cluster, we take the total
propagation length to be 2 Mpc, and for central AGNs we take a propagation length of 1 Mpc. For sources embedded inside a cluster (where the 3-dimensional
location is unknown) we also use a propagation length of 1 Mpc.\footnote{For a source towards the front (rear) of the a cluster, this will result in bounds
that are overly strong (weak).}
With these choices of lengths, we are effectively treating the linear size of the cluster as 2 Mpc.
The exact boundary is somewhat arbitrary, but as the magnetic field falls off with the distance from the centre of the cluster
(as $B(r) \propto n_e(r)^{\eta}$), conversion is suppressed far away from the cluster centre.

We draw magnetic field domain lengths according to a probability distribution
$$
{\rm Prob}(L) \propto L^{-1/3}.
$$
This power spectrum involves a range of scales, with power predominantly at larger scales.
It is therefore sensitive to the maximal length scale $L_{max}$ used in the model.
This index is based on the Kolmogorov power spectrum ${\rm Prob}(\vert k \vert) \propto k^{-5/3}$.

There are three important caveats to be placed on the magnetic fields used. The first is the uncertainty in the statistical characterisation of the magnetic field.
Existing radio observations are only able to do this for a limited number of cases (for example Coma \cite{Bonafede}). While one expects broad similarity in the magnitudes and coherence lengths of the magnetic field between clusters, this represents an uncertainty that will require future observations with the Square Kilometre Array to reduce.

The second caveat is that even if the statistical characterisation of the magnetic field were known exactly, it would still be the case
that the field profile along the actual line of sight to the source would be unknown.
Different field profiles may result in photon-ALP conversion probabilities that would be easier or harder to observe. This uncertainty is irreducible for a single source, but can be
reduced by considering multiple sources as in this paper, where the chance of multiple `bad' configurations becomes negligible.

The final caveat is that on physical grounds, one might expect the coherence lengths to be smaller towards the centre of clusters, and then
increase towards the edges where all the characteristic scales become larger (for example, see the theoretical analysis in \cite{Kunz}). This would be particularly
important for off-centre AGNs which are displaced from the cluster centre.
This effect is not taken into account in our model that uses a fixed maximal and minimal domain length.

\subsection{Methodology for ALP Constraints}
\label{methodology}

ALPs are constrained because they would lead to unobserved spectral modulations. For each source, we first fit an absorbed power law
\begin{equation}
A E^ {- \gamma} \times e^{-n_{H} \sigma(E)}\, ,
\label{eq:absorbedpowerlaw}
\end{equation}
 (possibly supplemented by a soft thermal component) to the actual data. For all the sources studied in this paper,
this default astrophysical model produces a good fit (in contrast to the study of the central Perseus AGN NGC1275 in \cite{160501043, 160801684}).
We therefore seek to set bounds on $g_{a\gamma\gamma}$ by determining the maximal level of the
ALP-photon coupling consistent with the data.

To do so, we need to determine the expected level of spectral modulations.
For each source, we generate 100 magnetic fields according to the parameters in Table~\ref{SourceMagneticFields}.
For each fixed magnetic field, we
use a range of photon-ALP couplings $g_{a\gamma\gamma}$ between $5 \ti 10^{-13} {\rm GeV}^{-1}$ and $10^{-11} {\rm GeV}^{-1}$,
and evaluate a table of photon-ALP conversion probabilities from 0 to 8 keV.
In the presence of the magnetic field, the `mass' eigenstate of the Hamiltonian is a mixture of photon and ALP eigenstates.
The conversion probabilities are determined by
propagating a photon `flavour' eigenstate through a series of magnetic field domains and evaluating its eventual overlap with the ALP `flavour' eigenstate.
This calculation is performed using the formalism originally described in \cite{RaffeltStodolsky} (a more recent treatment is e.g. \cite{13123947}).

Given a particular magnetic field realisation and coupling $g_{a\gamma\gamma}$, this defines a model for the
 spectral modulations expected in the presence of ALPs.
For each model, we generate ten fake data samples using the sherpa command {\tt fakepha}, using the same
exposure and responses as for the actual data. The fake data samples are generated using the best fit no-ALP model multiplied by the photon survival probability. This gives a
total of one thousand fake data samples, generated under the assumption that ALPs exist with a specified coupling (the null hypothesis).
For each source, we fit an absorbed power law to the fake data, using exactly the same binning and energy ranges as
for the real data. We allow the spectral index and amplitude to vary independently in each fit.

This gives rise to bounds on the
ALP-photon coupling, as in the case of a large coupling the expected (i.e. simulated) data
will contain large spectral modulations that would result in a bad fit and so be incompatible with the real data.
As the coupling is reduced, the size of these modulations will become smaller, leading to compatibility with the real data.

In terms of producing bounds, we adopt a frequentist perspective. We take the null hypothesis to be the existence of ALPs with a specified coupling,
and seek to exclude this hypothesis.
 We regard a coupling $g_{a\gamma\gamma}$ as excluded at
90\% confidence level when 90\% of the fake data samples lead to a worse fit (i.e. a higher reduced $\chi^2$, that we also require to be $> 1$) than for the
actual data.\footnote{It is important to note here that what we are really constraining is the combination of the coupling $g_{a \gamma \gamma}$ \emph{and}
the model of the magnetic field used. The uncertainty in the central magnetic field, for example, is hard to quantify in a formal manner, but probably involves a factor of two (note that for
the model of the Coma magnetic field in \cite{Bonafede}, the 1$\sigma$ constraint on the central magnetic field is $B_0 = 3.9 - 5.4 \mu {\rm G}$).
This source of uncertainty will require the SKA
to reduce it. Of course, this caveat also applies to all other work on constraining $g_{a\gamma\gamma}$ using astrophysical sources.} In the case that the no-ALP model fits the data with a reduced $\chi^2 < 1$, we still only consider fits with reduced $\chi^2 > 1$ to be a worse fit than for the actual data.

\section{Individual Sources, Spectra and Constraints}
\label{sec:indsources}

\subsection{Quasar B1256+281 behind Coma}

This quasar is located behind the Coma cluster at a redshift of $z=0.38$. Its sightline passes through the entirety of the Coma intra-cluster medium (ICM).
There are around 5000 counts from the source, of which around 10\% can be attributed to the ICM (as the source is always off-axis, the \emph{Chandra}
Point Spread Function is degraded
compared to an on-axis observation, increasing the level of contamination from thermal cluster emission).

\begin{figure}\includegraphics[width=1.0\textwidth]{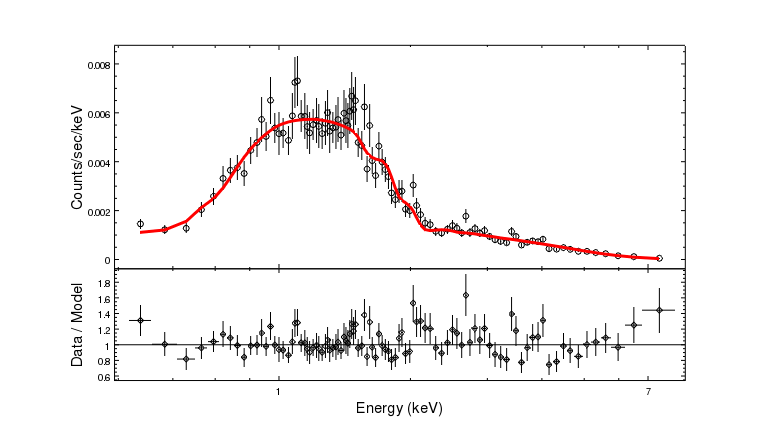}
\caption{The stacked spectrum of the quasar B1256+281 behind Coma. The fit is an unabsorbed power law of index $\gamma = (1.75 \pm 0.03)$.}
\label{ComaQuasarIfig}
\end{figure}

Grouping counts so that there are at least 40 counts per bin, the quasar spectrum from 0.5 to 7 keV is well-fit by an unabsorbed power-law with index $1.75 \pm 0.04$ (a reduced
$\chi^2$ of 0.88 for 96 degrees of freedom). There is no requirement for an Fe K$\alpha$ line. The spectrum is plotted in Figure~\ref{ComaQuasarIfig}. No significant residuals are observed, and we can say that there are no ALP-induced modulations in the spectrum beyond the 20\% level.

Simulating fake data with an ALP present in the spectrum as described in Section~\ref{methodology},
the 95\% confidence level bound is $g_{a\gamma\gamma} < 6 \ti 10^{-12} {\rm GeV}^{-1}$.

\subsection{Quasar SDSS J130001.47+275120.6 behind Coma}

This is an even more distant quasar, at redshift 0.975. The sightline again passes through the entirety of the Coma ICM.
There are around 3000 counts after background subtraction (around 3800 prior), and the resulting spectrum is well fit by the sum of an
unabsorbed power law with index $\gamma = (1.80 \pm 0.05)$ and an Fe K$\alpha$ line at 6.4 keV in the rest frame (inclusion of the Fe line gives
an improvement of $\Delta \chi^2 = 5$ in the fit, and so is 2.2$\sigma$ preferred).  Fitting from 0.5 to 7 keV, and
grouping counts so that there at least 30 counts per bin, the reduced $\chi^2$ is 0.99 for 95 degrees of freedom.

\begin{figure}\includegraphics[width=1.0\textwidth]{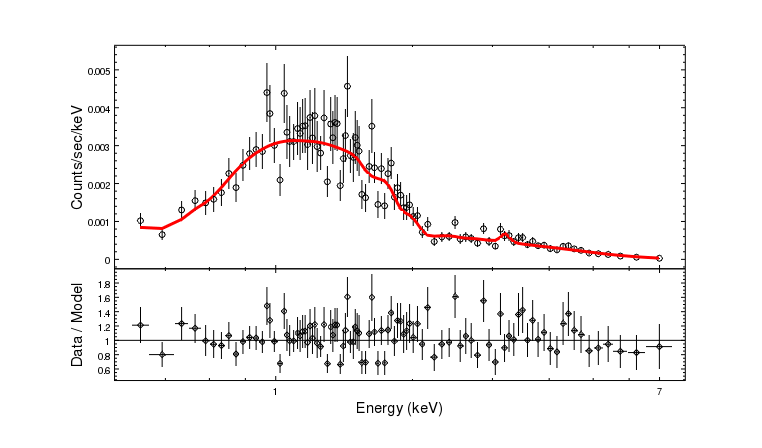}
\caption{The stacked spectrum of the $z=0.975$ quasar SDSS J130001.48+275120.6 behind Coma. The fit is an unabsorbed power law of index $\gamma = (1.80 \pm 0.05)$ plus
an Fe K$\alpha$ line at 6.4 keV in the rest frame.}
\label{ComaQuasarII}
\end{figure}

The spectrum is plotted in Figure~\ref{ComaQuasarII}.
The relatively small number of counts means that we can only restrict ALP-induced spectral irregularities to the $\lesssim 30\%$ level, as there is not the statistical leverage to
constrain beyond that.

The small counts means that it is not possible to place a 95\% confidence level exclusion. (A value of $g_{a\gamma\gamma} = 10^{-11} {\rm GeV}^{-1}$ is excluded
at 87\% confidence level.)

\subsection{NGC3862 within A1367}

\begin{figure}\includegraphics[width=1.0\textwidth]{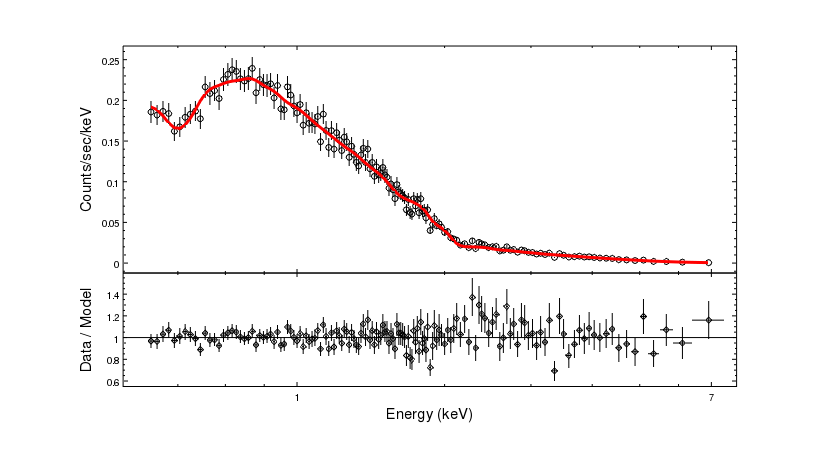}
\caption{The stacked spectrum of the AGN NGC3862 in A1367. The fit is the sum of a  power law of index $\gamma = (2.30 \pm 0.05)$ plus
a soft thermal component with temperature 0.3 keV, absorbed by a column density of $n_H = 5 \ti 10^{20} {\rm cm}^{-2}$.}
\label{NGC3862}
\end{figure}

The AGN NGC3862 within the cluster A1367 is characterised by a very soft power-law (index $2.30 \pm 0.03$) absorbed by a column density of $n_H = 5 \ti 10^{20} {\rm cm}^{-2}$,
supplemented by a soft thermal component $T \sim 0.3 {\rm keV}$). Grouping counts so that there are at least 50 counts per bin, the reduced $\chi^2$ is 0.83 for
144 degrees of freedom, with a total of 21000 counts after background subtraction.
The spectrum is plotted in Figure~\ref{NGC3862} and the resulting fit shows no sign of any significant spectral irregularities.

The low electron density within A1367 increases the efficiency of ALP-photon conversion (as it reduces the effective mass differential between the photon and the ALP).
The large number of counts then allows good bounds to be obtained, $g_{a\gamma\gamma} < 2.4 \ti 10^{-12} {\rm GeV}^{-1}$ at 95\% confidence and $g_{a\gamma\gamma} < 2.9 \ti 10^{-12} {\rm GeV}^{-1}$ at 99\% confidence.

\subsection{Central AGN IC4374 of A3581}

\begin{figure}\includegraphics[width=1.0\textwidth]{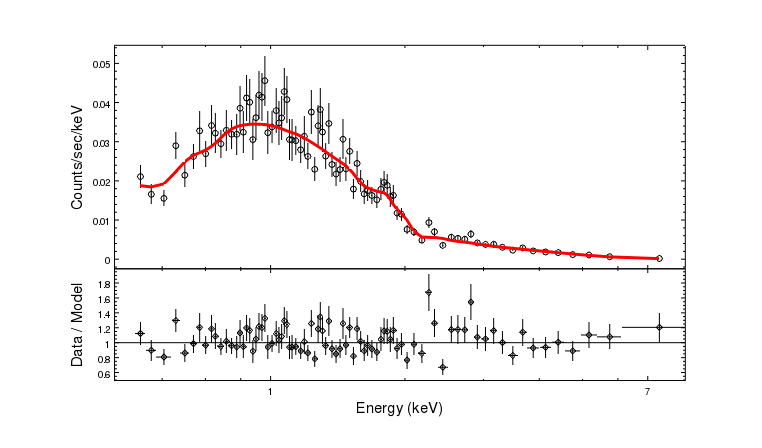}
\caption{The stacked spectrum of the central cluster galaxy IC4374 located in A3581. A reasonably good fit is provided by
an absorbed power law with $\gamma = 2.00 \pm 0.05$ and $n_H = (9 \pm 1.5) \ti 10^{20} {\rm cm}^{-2}$.
Grouping counts with at least 40 counts per bin, the reduced $\chi^2$ is 1.13 for 82 degrees of freedom.}
\label{A3581}
\end{figure}

This central AGN has around 4400 counts after background subtraction (4600 prior to background subtraction). A reasonably good fit is provided by
an absorbed power law with $\gamma = 2.00 \pm 0.05$ and $n_H = (9 \pm 1.5) \ti 10^{20} {\rm cm}^{-2}$.
Grouping counts with at least 40 counts per bin, the reduced $\chi^2$ is 1.13 for 82 degrees of freedom.
The spectrum is plotted in Figure~\ref{A3581}. As this is a very poor cluster, the central magnetic field is expected to be very weak.
This reduces conversion efficiency, with the result that this source is unable to provide any constraints - a combination of the small number of counts
and the weak magnetic field.



\subsection{Seyfert galaxy 2E3140 in A1795}

This is a bright unobscured AGN. Its redshift is 0.059, compared to a cluster redshift of 0.062. The radial velocity difference is $1000 \, {\rm km s}^{-1}$,
which is within the range of the A1795 velocity dispersion, and is consistent with 2E3140 being a bound member of the cluster A1795, with a sightline
that passes through the intracluster medium.

\begin{figure}\includegraphics[width=1.0\textwidth]{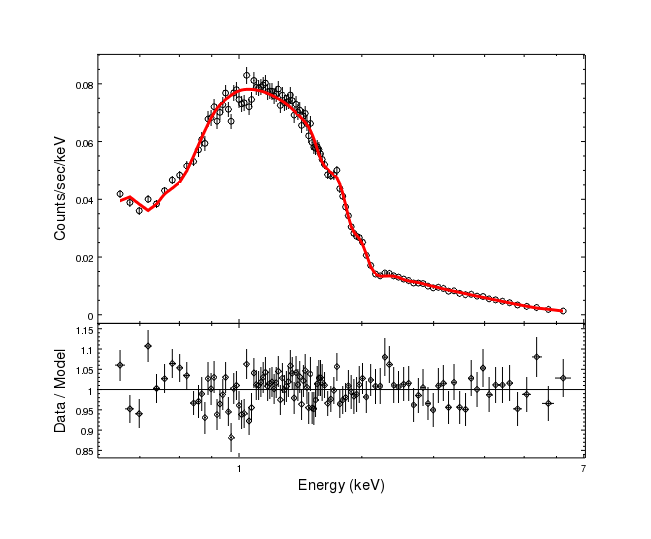}
\caption{The stacked spectrum of the bright Seyfert 1 galaxy 2E3140 located in A1795. The fit is the sum of a power-law with $\gamma = 2.11 \pm 0.01$, a soft thermal component with $T \sim 0.1 {\rm keV}$ and a weak
Fe K$\alpha$ line at 6.4 keV in the rest frame (${\tt powlaw1d + xsapec + xszgauss}$).
Grouping counts so that there are at leat 500 in each bin,
the overall fit is excellent with a reduced $\chi^2$ of 0.98 for 103 degrees of freedom.
}
\label{Sy1A1795}
\end{figure}

However, we do not know the precise 3D location within the intracluster medium, and therefore whether the line of sight
passes through most or only small amounts of the ICM. We assume a midway position.\footnote{This is perhaps supported by the large velocity relative to the cluster centre, as
 an object undergoing harmonic motion about a central source has maximal relative velocity at the midpoint of its oscillation.}
 The extracted spectrum contains around 78000 counts (of which around 1000 are ICM background).
The spectrum from 0.5 to 6 keV is very well fit by the sum of a power-law with index $\gamma = 2.11 \pm 0.01$, a soft thermal component with $T \sim 0.1 {\rm keV}$ and a weak
Fe K$\alpha$ line at 6.4 keV in the rest frame (the Sherpa model ${\tt powlaw1d + xsapec + xszgauss}$).
Consistent with the galactic $n_H$ column density, no absorption is required in the fit.
Grouping counts so that there are at least 500 in each bin,
the overall fit is excellent with a reduced $\chi^2$ of 0.98 for 103 degrees of freedom.
The spectrum is plotted in Figure~\ref{Sy1A1795}.

The large number of counts joined to the excellent fit results in strong bounds. Simulating fake data as in Section~\ref{methodology}, we
obtain $g_{a\gamma\gamma} < 1.5 \ti 10^{-12} {\rm GeV}^{-1}$ at 95\% confidence level and
$g_{a\gamma\gamma} < 1.6 \ti 10^{-12} {\rm GeV}^{-1}$ at 99\% confidence level.

\subsection{Quasar CXOUJ134905.8+263752 behind A1795}

\begin{figure}\includegraphics[width=1.0\textwidth]{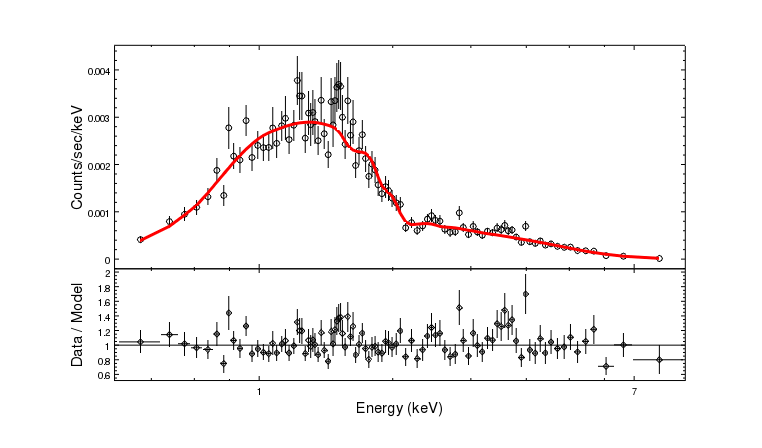}
\caption{The stacked spectrum of the $z=1.3$ quasar CXOU J134905.8+263752 behind A1795. The fit is to an absorbed power-law, with a local
absorption column density of $n_H = (1.0 \pm 0.1) \ti 10^{22} {\rm cm^{-2}}$, and a power-law index of
$\gamma = 1.61 \pm 0.04$. Bins are grouped so that each bin has at least 40 counts. The overall fit between 0.5 and 7 keV is good, with a reduced
$\chi^2$ of 1.12 for 96 degrees of freedom.
}
\label{A1795QuasarFig}
\end{figure}

This $z=1.3$ quasar is an obscured AGN with a total of around 5000 counts (5300 before background subtraction) arising from 985ks of \emph{Chandra} observation time.
It is fit by an absorbed power-law, with a local absorption column density of $n_H = (1.0 \pm 0.1) \ti 10^{22} {\rm cm^{-2}}$, and a power-law index of
$\gamma = 1.61 \pm 0.04$. No contribution from Milky Way absorption is required in the fit. Bins are grouped so that each bin has at least 40 counts. The overall fit between 0.5 and 7 keV is good, with a reduced $\chi^2$ of 1.12 for 96 degrees of freedom. The spectrum is shown in Figure~\ref{A1795QuasarFig}.

The relatively small number of counts means that no strong bounds
can be extracted for this source ($g_{a\gamma\gamma} < 10^{-11} {\rm GeV}^{-1}$ only at 75\% confidence level).

\subsection{UGC9799 in A2052}

\begin{figure}\includegraphics[width=1.0\textwidth]{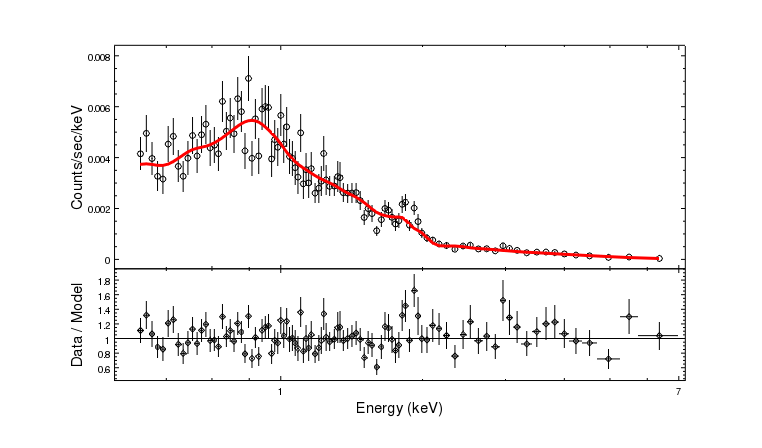}
\caption{The stacked spectrum of the central cluster galaxy UGC9799 located in A2052. The fit is to an unabsorbed power law with
index $\gamma = 1.85 \pm 0.04$ supplemented by a soft thermal component with $T \sim 0.9 \pm 0.2 {\rm keV}$. There are 4300 counts after background subtraction.
Fitting between 0.5 and 7 keV, and grouping counts so that there are at least 30 counts per bin, the reduced $\chi^2$ is 1.12 for 94 degrees of freedom.
}
\label{A2052}
\end{figure}

This central AGN of the cluster A2052 is well characterised by an unabsorbed power-law with
index $\gamma = 1.85 \pm 0.04$ supplemented by a soft thermal component with $T \sim 0.9 \pm 0.2 {\rm keV}$. There are 4300 counts after background subtraction.
Fitting between 0.5 and 7 keV, and grouping counts so that there are at least 30 counts per bin, the reduced $\chi^2$ is 1.12 for 94 degrees of freedom.
The soft spectrum, combined with the relatively low number of counts, implies that ALP-induced modulations are only excluded above the 20\% level.
The spectrum is plotted in Figure~\ref{A2052}.

As with A3581, the relatively small number of counts means that this source is unable to provide
useful constraints, and $g_{a\gamma\gamma} = 10^{-11} {\rm GeV}^{-1}$ is not excluded by this source.

\section{Conclusions and Outlook for the Future of X-Ray Axion Searches}
\label{sec:conclusions}

The search for spectral modulations from X-ray point sources represents a competitive alternative to lab-based experiments, such as IAXO, seeking to constrain or discover axions. These astrophysical searches are particularly strong in the low-mass regime $m_a \lesssim 10^{-12} {\rm eV}$
where the ALP mass is below the plasma frequency of galaxy clusters.


In this paper we have extended previous analyses using Hydra A \cite{WoutersBrun}, NGC1275 \cite{160501043} and M87 \cite{170307354} to include many new point sources.
No evidence for any ALP-induced spectral modulations has been found, and all the spectra are
consistent with absorbed power laws. While the sources here are not (with the exception of 2E3140) as individually constraining as either NGC1275 or M87, they sample many more
sightlines, which is necessary to minimise the risk of having a single `bad' magnetic field along the sightline to a single source.

This papers also contains the first results using displaced cluster sources, which are not at the dynamic centre of the cluster.
The advantage of such displaced sources it that they samples a different environment than those located at cluster centres.
In particular, for displaced sources the electron density is smaller and the magnetic field coherence lengths are expected to be larger.
Given the intrinsic uncertainties on magnetic fields for any one single source, in developing constraints on ALPs it is important to build up a diverse sample
of many different objects. We have also found two bright displaced sources (2E3140 and NGC3862) where the constraints on ALPs are comparable to those from NGC1275 and M87.

There are two directions in which current results can be improved. The first involves a more accurate knowledge of cluster magnetic fields.
As ALP searches are sensitive to the product $g_{a\gamma\gamma} B_{\perp}$, the constraint on $g_{a\gamma\gamma}$ is only as good as the knowledge of
$B_{\perp}$. The most accurate knowledge of cluster magnetic fields is obtained through Faraday Rotation measures of many individual sources either shining through
or embedded in galaxy clusters.
By allowing access to weaker sources, the advent of the SKA will greatly improve knowledge of cluster magnetic fields, and will be particularly useful for constraining the
Perseus and Virgo magnetic fields.

The second involves the quality of X-ray telescope data.
Current searches for new physics with X-ray data rely on the accumulated data archives of either \emph{Chandra} (which offers the best angular resolution)
or \emph{XMM-Newton} (which has the highest effective area). Significant improvements will be attained with the launch of
ATHENA, which will combine excellent angular resolution with a great increase in effective area and the use of microcalorimeters
for dramatically improved energy resolution. In particular, ATHENA will allow sensitivity to the rapid
oscillatory structure in ALP-photon conversion below 2 keV. Before ATHENA's launch in 2028, further benefits may arise from the \emph{Hitomi} recovery mission XARM
or the polarisation experiments IXPE \cite{IXPE} or eXTP.

The combination of advances in magnetic field knowledge and advances in the quality of X-ray data should lead to further improvements in the ability to constrain the ALP-photon
coupling over the next decade.

\section*{Acknowledgments}

We thank David Marsh for helpful comments on the draft.
This research has made use of the NASA/IPAC Extragalactic Database (NED) which is operated by the Jet Propulsion Laboratory, California Institute of Technology, under contract with the National Aeronautics and Space Administration.
This research has made use of data obtained from the Chandra Data Archive and the Chandra Source Catalog, and software provided by the Chandra X-ray Center (CXC) in the application packages CIAO, ChIPS, and Sherpa.
This project is funded in part by the European Research Council starting grant `Supersymmetry Breaking in String Theory' (307605).

\bibliography{refs}

\providecommand{\href}[2]{#2}\begingroup\raggedright\begin{thebibliography}{10}

\bibitem{hepth0602233}
J.~P. Conlon, {\it {The QCD axion and moduli stabilisation}},  {\em JHEP} {\bf
  05} (2006) 078, [\href{http://xxx.lanl.gov/abs/hep-th/0602233}{{\tt
  hep-th/0602233}}].

\bibitem{hepth0605206}
P.~Svrcek and E.~Witten, {\it {Axions In String Theory}},  {\em JHEP} {\bf 06}
  (2006) 051, [\href{http://xxx.lanl.gov/abs/hep-th/0605206}{{\tt
  hep-th/0605206}}].

\bibitem{1206.0819}
M.~Cicoli, M.~Goodsell, and A.~Ringwald, {\it {The type IIB string axiverse and
  its low-energy phenomenology}},  {\em JHEP} {\bf 10} (2012) 146,
  [\href{http://xxx.lanl.gov/abs/1206.0819}{{\tt 1206.0819}}].

\bibitem{RingwaldReview}
A.~Ringwald, {\it {Exploring the Role of Axions and Other WISPs in the Dark
  Universe}},  {\em Phys.Dark Univ.} {\bf 1} (2012) 116--135,
  [\href{http://xxx.lanl.gov/abs/1210.5081}{{\tt 1210.5081}}].

\bibitem{Sikivie:1983ip}
P.~Sikivie, {\it {Experimental Tests of the Invisible Axion}},  {\em Phys. Rev.
  Lett.} {\bf 51} (1983) 1415--1417. [Erratum: Phys. Rev. Lett.52,695(1984)].

\bibitem{Raffelt}
G.~Raffelt and L.~Stodolsky, {\it {Mixing of the Photon with Low Mass
  Particles}},  {\em Phys.Rev.} {\bf D37} (1988) 1237.

\bibitem{IAXO}
{\bf IAXO} Collaboration, I.~Irastorza {\em et.~al.}, {\it {The International
  Axion Observatory IAXO. Letter of Intent to the CERN SPS committee}}, .

\bibitem{Brockway:1996yr}
J.~W. Brockway, E.~D. Carlson, and G.~G. Raffelt, {\it {SN1987A gamma-ray
  limits on the conversion of pseudoscalars}},  {\em Phys. Lett.} {\bf B383}
  (1996) 439--443, [\href{http://xxx.lanl.gov/abs/astro-ph/9605197}{{\tt
  astro-ph/9605197}}].

\bibitem{Grifols:1996id}
J.~A. Grifols, E.~Masso, and R.~Toldra, {\it {Gamma-rays from SN1987A due to
  pseudoscalar conversion}},  {\em Phys. Rev. Lett.} {\bf 77} (1996)
  2372--2375, [\href{http://xxx.lanl.gov/abs/astro-ph/9606028}{{\tt
  astro-ph/9606028}}].

\bibitem{1410.3747}
A.~Payez, C.~Evoli, T.~Fischer, M.~Giannotti, A.~Mirizzi, and A.~Ringwald, {\it
  {Revisiting the SN1987A gamma-ray limit on ultralight axion-like particles}},
   {\em JCAP} {\bf 1502} (2015), no.~02 006,
  [\href{http://xxx.lanl.gov/abs/1410.3747}{{\tt 1410.3747}}].

\bibitem{09014085}
M.~Fairbairn, T.~Rashba, and S.~V. Troitsky, {\it {Photon-axion mixing and
  ultra-high-energy cosmic rays from BL Lac type objects - Shining light
  through the Universe}},  {\em Phys. Rev.} {\bf D84} (2011) 125019,
  [\href{http://xxx.lanl.gov/abs/0901.4085}{{\tt 0901.4085}}].

\bibitem{0902.2320}
C.~Burrage, A.-C. Davis, and D.~J. Shaw, {\it {Active Galactic Nuclei Shed
  Light on Axion-like-Particles}},  {\em Phys. Rev. Lett.} {\bf 102} (2009)
  201101, [\href{http://xxx.lanl.gov/abs/0902.2320}{{\tt 0902.2320}}].

\bibitem{WoutersBrun}
D.~Wouters and P.~Brun, {\it {Constraints on Axion-like Particles from X-Ray
  Observations of the Hydra Galaxy Cluster}},  {\em Astrophys. J.} {\bf 772}
  (2013) 44, [\href{http://xxx.lanl.gov/abs/1304.0989}{{\tt 1304.0989}}].

\bibitem{1305.3603}
J.~P. Conlon and M.~C.~D. Marsh, {\it {Excess Astrophysical Photons from a
  0.1-1 keV Cosmic Axion Background}},  {\em Phys. Rev. Lett.} {\bf 111}
  (2013), no.~15 151301, [\href{http://xxx.lanl.gov/abs/1305.3603}{{\tt
  1305.3603}}].

\bibitem{1312.3947}
S.~Angus, J.~P. Conlon, M.~C.~D. Marsh, A.~J. Powell, and L.~T. Witkowski, {\it
  {Soft X-ray Excess in the Coma Cluster from a Cosmic Axion Background}},
  {\em JCAP} {\bf 1409} (2014), no.~09 026,
  [\href{http://xxx.lanl.gov/abs/1312.3947}{{\tt 1312.3947}}].

\bibitem{150101642}
G.~D'Amico and N.~Kaloper, {\it {Anisotropies in nonthermal distortions of
  cosmic light from photon-axion conversion}},  {\em Phys. Rev.} {\bf D91}
  (2015), no.~8 085015, [\href{http://xxx.lanl.gov/abs/1501.01642}{{\tt
  1501.01642}}].

\bibitem{150702855}
M.~Schlederer and G.~Sigl, {\it {Constraining ALP-photon coupling using galaxy
  clusters}},  {\em JCAP} {\bf 1601} (2016), no.~01 038,
  [\href{http://xxx.lanl.gov/abs/1507.02855}{{\tt 1507.02855}}].

\bibitem{1509.06748}
J.~P. Conlon, M.~C.~D. Marsh, and A.~J. Powell, {\it {Galaxy Cluster Thermal
  X-Ray Spectra Constrain Axion-Like Particles}},
  \href{http://xxx.lanl.gov/abs/1509.06748}{{\tt 1509.06748}}.

\bibitem{12046187}
A.~Payez, J.~R. Cudell, and D.~Hutsemekers, {\it {New polarimetric constraints
  on axion-like particles}},  {\em JCAP} {\bf 1207} (2012) 041,
  [\href{http://xxx.lanl.gov/abs/1204.6187}{{\tt 1204.6187}}].

\bibitem{150906748}
J.~P. Conlon, M.~C.~D. Marsh, and A.~J. Powell, {\it {Galaxy cluster thermal
  x-ray spectra constrain axionlike particles}},  {\em Phys. Rev.} {\bf D93}
  (2016), no.~12 123526, [\href{http://xxx.lanl.gov/abs/1509.06748}{{\tt
  1509.06748}}].

\bibitem{161205697}
Y.~Gong, X.~Chen, and H.~Feng, {\it {Testing the axion-conversion hypothesis of
  3.5 keV emission with polarization}},  {\em Phys. Rev. Lett.} {\bf 118}
  (2017), no.~6 061101, [\href{http://xxx.lanl.gov/abs/1612.05697}{{\tt
  1612.05697}}].

\bibitem{Profumo:2006im}
S.~Profumo and K.~Sigurdson, {\it {The Shadow of Dark Matter}},  {\em Phys.
  Rev.} {\bf D75} (2007) 023521,
  [\href{http://xxx.lanl.gov/abs/astro-ph/0611129}{{\tt astro-ph/0611129}}].

\bibitem{160801684}
J.~P. Conlon, F.~Day, N.~Jennings, S.~Krippendorf, and M.~Rummel, {\it
  {Consistency of Hitomi, XMM-Newton and Chandra 3.5 keV data from Perseus}},
  \href{http://xxx.lanl.gov/abs/1608.01684}{{\tt 1608.01684}}.

\bibitem{1205.6428}
D.~Wouters and P.~Brun, {\it {Irregularity in gamma ray source spectra as a
  signature of axionlike particles}},  {\em Phys. Rev.} {\bf D86} (2012)
  043005, [\href{http://xxx.lanl.gov/abs/1205.6428}{{\tt 1205.6428}}].

\bibitem{BraxWoutersBrun}
P.~Brax, P.~Brun, and D.~Wouters, {\it {Galaxy cluster constraints on the
  coupling to photons of low-mass scalars}},  {\em Phys. Rev.} {\bf D92} (2015)
  083501, [\href{http://xxx.lanl.gov/abs/1505.01020}{{\tt 1505.01020}}].

\bibitem{1603.06978}
{\bf Fermi-LAT} Collaboration, M.~Ajello {\em et.~al.}, {\it {Search for
  Spectral Irregularities due to Photon–Axionlike-Particle Oscillations with
  the Fermi Large Area Telescope}},  {\em Phys. Rev. Lett.} {\bf 116} (2016),
  no.~16 161101, [\href{http://xxx.lanl.gov/abs/1603.06978}{{\tt 1603.06978}}].

\bibitem{160501043}
M.~Berg, J.~P. Conlon, F.~Day, N.~Jennings, S.~Krippendorf, A.~J. Powell, and
  M.~Rummel, {\it {Searches for Axion-Like Particles with NGC1275: Observation
  of Spectral Modulations}},  \href{http://xxx.lanl.gov/abs/1605.01043}{{\tt
  1605.01043}}.

\bibitem{Bulbul}
E.~Bulbul, M.~Markevitch, A.~Foster, R.~K. Smith, M.~Loewenstein, {\em
  et.~al.}, {\it {Detection of An Unidentified Emission Line in the Stacked
  X-ray spectrum of Galaxy Clusters}},  {\em Astrophys.J.} {\bf 789} (2014) 13,
  [\href{http://xxx.lanl.gov/abs/1402.2301}{{\tt 1402.2301}}].

\bibitem{Boyarsky}
A.~Boyarsky, O.~Ruchayskiy, D.~Iakubovskyi, and J.~Franse, {\it {An
  unidentified line in X-ray spectra of the Andromeda galaxy and Perseus galaxy
  cluster}},  \href{http://xxx.lanl.gov/abs/1402.4119}{{\tt 1402.4119}}.

\bibitem{170307354}
M.~C.~D. Marsh, H.~R. Russell, A.~C. Fabian, B.~P. McNamara, P.~Nulsen, and
  C.~S. Reynolds, {\it {A New Bound on Axion-Like Particles}},
  \href{http://xxx.lanl.gov/abs/1703.07354}{{\tt 1703.07354}}.

\bibitem{SIMBAD}
M.~Wenger {\em et.~al.}, {\it {The simbad astronomical database}},  {\em
  Astron. Astrophys. Suppl. Ser.} {\bf 143} (2000) 9,
  [\href{http://xxx.lanl.gov/abs/astro-ph/0002110}{{\tt astro-ph/0002110}}].

\bibitem{Bonafede}
A.~Bonafede, L.~Feretti, M.~Murgia, F.~Govoni, G.~Giovannini, D.~Dallacasa,
  K.~Dolag, and G.~B. Taylor, {\it {The Coma cluster magnetic field from
  Faraday rotation measures}},  {\em Astron. Astrophys.} {\bf 513} (2010) A30,
  [\href{http://xxx.lanl.gov/abs/1002.0594}{{\tt 1002.0594}}].

\bibitem{12014119}
V.~Vacca, M.~Murgia, F.~Govoni, L.~Feretti, G.~Giovannini, R.~A. Perley, and
  G.~B. Taylor, {\it {The intracluster magnetic field power spectrum in
  A2199}},  {\em Astron. Astrophys.} {\bf 540} (2012) A38,
  [\href{http://xxx.lanl.gov/abs/1201.4119}{{\tt 1201.4119}}].

\bibitem{A194}
F.~Govoni {\em et.~al.}, {\it {Sardinia Radio Telescope observations of Abell
  194 - the intra-cluster magnetic field power spectrum}},
  \href{http://xxx.lanl.gov/abs/1703.08688}{{\tt 1703.08688}}.

\bibitem{KucharEnsslin}
P.~{Kuchar} and T.~A. {En{\ss}lin}, {\it {Magnetic power spectra from Faraday
  rotation maps. REALMAF and its use on Hydra A}},  {\em Astron.Astrophys.}
  {\bf 529} (May, 2011) A13.

\bibitem{Ensslin97}
T.~A. Ensslin, P.~L. Biermann, U.~Klein, and S.~Kohle, {\it {Cluster radio
  relics as a tracer of shock waves of the large - scale structure formation}},
   {\em Astron. Astrophys.} {\bf 332} (1998) 395,
  [\href{http://xxx.lanl.gov/abs/astro-ph/9712293}{{\tt astro-ph/9712293}}].

\bibitem{Henriksen}
H.~V. {Klapdor-Kleingrothaus} and I.~V. {Krivosheina}, eds., {\em {Dark Matter
  in Astrophysics and Particle Physics - Proceedings of the 7th International
  Heidelberg Conference on Dark 2009}}, 2010.

\bibitem{GeOwen1993}
J.~P. {Ge} and F.~N. {Owen}, {\it {Faraday rotation in cooling flow clusters of
  galaxies. I - Radio and X-ray observations of Abell 1795}},  {\em ApJ} {\bf
  105} (Mar., 1993) 778--787.

\bibitem{Ettori2000}
S.~{Ettori}, {\it {{$\beta$}-model and cooling flows in X-ray clusters of
  galaxies}},  {\em MNRAS} {\bf 318} (Nov., 2000) 1041--1046,
  [\href{http://xxx.lanl.gov/abs/astro-ph/0005224}{{\tt astro-ph/0005224}}].

\bibitem{ChandranDennis2005}
T.~J. {Dennis} and B.~D.~G. {Chandran}, {\it {Turbulent Heating of
  Galaxy-Cluster Plasmas}},  {\em ApJ} {\bf 622} (Mar., 2005) 205--216.

\bibitem{Blanton2003}
E.~L. {Blanton}, C.~L. {Sarazin}, and B.~R. {McNamara}, {\it {Chandra
  Observation of the Cooling Flow Cluster Abell 2052}},  {\em ApJ} {\bf 585}
  (Mar., 2003) 227--243, [\href{http://xxx.lanl.gov/abs/astro-ph/0211027}{{\tt
  astro-ph/0211027}}].

\bibitem{Machado2014}
R.~E.~G. {Machado} and G.~B. {Lima Neto}, {\it {Simulations of gas sloshing in
  galaxy cluster Abell 2052}},  {\em MNRAS} {\bf 447} (Mar., 2015) 2915--2924,
  [\href{http://xxx.lanl.gov/abs/1412.4558}{{\tt 1412.4558}}].

\bibitem{Johnstone98}
R.~M. {Johnstone}, A.~C. {Fabian}, and G.~B. {Taylor}, {\it {X-ray and radio
  observations of the poor cluster A3581 which hosts the radio galaxy PKS
  1404-267}},  {\em MNRAS} {\bf 298} (Aug., 1998) 854--860.

\bibitem{Johnstone2005}
R.~M. {Johnstone}, A.~C. {Fabian}, R.~G. {Morris}, and G.~B. {Taylor}, {\it
  {The galaxy cluster Abell 3581 as seen by Chandra}},  {\em MNRAS} {\bf 356}
  (Jan., 2005) 237--246, [\href{http://xxx.lanl.gov/abs/astro-ph/0410154}{{\tt
  astro-ph/0410154}}].

\bibitem{Kunz}
M.~W. Kunz, A.~A. Schekochihin, S.~C. Cowley, J.~J. Binney, and J.~S. Sanders,
  {\it {A thermally stable heating mechanism for the intracluster medium:
  turbulence, magnetic fields and plasma instabilities}},  {\em Mon. Not. Roy.
  Astron. Soc.} {\bf 410} (2011) 2446,
  [\href{http://xxx.lanl.gov/abs/1003.2719}{{\tt 1003.2719}}].

\bibitem{RaffeltStodolsky}
G.~Raffelt and L.~Stodolsky, {\it {Mixing of the Photon with Low Mass
  Particles}},  {\em Phys.Rev.} {\bf D37} (1988) 1237.

\bibitem{13123947}
S.~Angus, J.~P. Conlon, M.~C.~D. Marsh, A.~Powell, and L.~T. Witkowski, {\it
  {Soft X-ray Excess in the Coma Cluster from a Cosmic Axion Background}},
  \href{http://xxx.lanl.gov/abs/1312.3947}{{\tt 1312.3947}}.

\bibitem{IXPE}
M.~C. {Weisskopf}, B.~{Ramsey}, S.~{O'Dell}, A.~{Tennant}, R.~{Elsner},
  P.~{Soffitta}, R.~{Bellazzini}, E.~{Costa}, J.~{Kolodziejczak}, V.~{Kaspi},
  F.~{Muleri}, H.~{Marshall}, G.~{Matt}, and R.~{Romani}, {\it {The Imaging
  X-ray Polarimetry Explorer (IXPE)}},  in {\em Space Telescopes and
  Instrumentation 2016: Ultraviolet to Gamma Ray}, vol.~9905 of {\em SPIE
  Proceedings}, p.~990517, July, 2016.

\end{thebibliography}\endgroup
\bibliographystyle{JHEP}

\end{document}